# A heterojunction modulation-doped Mott transistor

Junwoo Son[1], Siddharth Rajan[2], Susanne Stemmer[1], and S. James Allen[3]

[1] Materials Department, University of California, Santa Barbara, California 93106-5050, U.S.A.

[2] Department of Electrical and Computer Engineering, The Ohio State University, Columbus, OH 43210 USA

[3] Department of Physics, University of California, Santa Barbara, CA 93106-9530, U.S.A.

**Abstract**

A heterojunction Mott field effect transistor (FET) is proposed that consists of an epitaxial channel material that exhibits an electron correlation induced Mott metal-to-insulator transition. The Mott material is remotely (modulation) doped with a degenerately doped conventional band insulator. An applied voltage modulates the electron transfer from the doped band insulator to the Mott material and produces transistor action by inducing an insulator-to-metal transition. Materials parameters from rare-earth nickelates and $SrTiO_3$ are used to assess the potential of the "modulation-doped Mott FET" (ModMottFET or MMFET) as a next-generation switch. It is shown that the MMFET is characterized by unique "charge gain" characteristics as well as competitive transconductance, small signal gain and current drive.

Electrostatic control of the competing electronic phases of transition metal oxides ("Mott materials") that are close to a Mott metal-insulator transition may open an important arena for future electronics [1,2] as well as provide an important tool for exploring the materials physics of highly correlated electron systems [3]. There have been a number of proposals for field effect transistors (FETs) based on control of the Mott transition [1,2] and some demonstrations of electrostatic gating of Mott materials [4]. The potential advantages of a Mott FET have been pointed out – large on/off ratios and femtosecond switching speeds [2]. An ideal Mott insulator is characterized by a half-filled $d$-band, corresponding to approximately one unpaired electron per unit cell, with the insulating ground state imposed by strong electron-electron repulsion [5]. The insulating state is upset if extra electrons or holes are introduced, which move the system away from a precisely half-filled $d$-band [5,6]. Typical estimates assume that complete suppression of the Mott insulating phase by electric-field-controlled electron accumulation or depletion requires very large charge carrier modulations, on the order of 0.1-0.3 carriers per unit cell [3]. Electrolyte gating can introduce carrier densities of this order and has been used to substantially alter the electronic state of Mott materials like $NdNiO_3$ [7,8] or to induce superconductivity in $SrTiO_3$ [9]. While electrolyte gating can induce a very large surface charge and is a powerful tool to explore the material physics of Mott material surfaces, it seems unlikely to be a viable approach for future electronics because the switching speed is limited by the low mobility of ions in the electrolytes.

In this letter we explore the performance of a Mott heterojunction field effect transistor that features a Mott insulator/band insulator interface and remote (modulation) doping of the Mott insulator. Modulation doping is widely used in high-mobility III-V semiconductor devices [10] and has recently been applied to Mott insulators such as Sm$_2$CuO$_4$ and LaVO$_3$ [11,12]. Introduction of modest space charge through charge transfer across the heterojunction into the Mott insulator will drive the Mott channel into the conducting state. A voltage applied across the interface controls the electron transfer. We will refer to the proposed FET as a "modulation-doped Mott FET", or ModMottFET (MMFET). By judicious choice of operating temperature and, most importantly, the Mott material, the model suggests that only ~ 0.01-0.02 electrons per unit cell can release ~ 1 electron per unit cell as the device is turned on, an effective charge gain with implications for speed and transconductance. We will first discuss the concept of charge gain followed by a discussion of the MMFET performance using specific materials parameters.

Like other Mott FETs [2], but unlike any other 3-terminal FET, the MMFET is a unique "charge gain" device. Gating only 1-10% of the channel electrons unleashes the entire conduction band of electrons in the Mott material. This is a completely new paradigm for device technology and in principle could lead to a pathway towards low power dissipation and ultra-fast devices as Si based transistors reach their limits [13]. To illustrate the concept of charge gain, we compare the operation of the Mott FET with a traditional FET structure using a simple circuit model of a traditional FET structure and that of a MMFET (see Supporting Information). The voltage applied between the gate and source causes *equal and opposite* charges to be induced in the gate and the channel, with the channel charge participating in conduction. Transistor gain can be explained in the small signal approximation as follows: a small voltage $\Delta v$ causes a change in charge, $q\Delta n = C_{gs}\Delta v$, which then changes the current. A *current* gain is achieved whenever the displacement current required to charge the gate-source capacitance ($C_{gs}$) is lower than the induced channel current.

In the case of a Mott FET, the electrostatic charge induced in the gate and channel are still equal (for charge neutrality). However, if a change of electron concentration leads to a Mott metal-insulator transition, then the electron density available for conduction may rise (or fall) by a few orders of magnitude. The total change in the *conducting charge* can then be much greater than the charge imaged on the gate. While the magnitude depends on the details of the Mott transition induced in the material (see below) it can be up to 100 times greater than the original charge induced by the gate bias. Since the charge imaged at the gate is essentially "amplified" into a higher conducting carrier density, we term this phenomenon *charge gain*. Assuming that the induced channel charge Δ$n$ is $K$ times the charge imaged at the gate Δ$n_G$, the frequency for unity current gain (cutoff frequency $f_\tau$) of such a transistor is given by (see Supporting Information):

$$f_\tau = K \frac{v_{eff}}{2\pi l}, \tag{1}$$

where $v_{eff}$ is the effective velocity of electrons in the channel, and $l$ is the gate length. In the Mott FET, the cutoff frequency is therefore *enhanced* by a factor of $K$ over the cutoff frequency of a purely capacitive FET.

For the purpose of discussing the performance of the MMFET, we use the known effect of *chemical doping* on the phase transition and conductivity of a rare earth nickelate, NdNiO$_3$. Rare earth nickelates ($R$NiO$_3$, where $R$ is a trivalent rare earth ion but not La) have been extensively studied in the bulk as they exhibit a sharp first order phase transition to an insulating state upon cooling [14]. The transition temperature can be lowered by chemical doping [15,16]. After accounting for lattice distortions that are also introduced by chemical substitution, the authors in ref. [15] conclude that a 1% change in electron or hole concentration suppresses the transition temperature by ~ 25 to 50 °K for electrons or holes, respectively. Nevertheless, the pathways for the metal-to-insulator transition of the rare earth nickelates and for the origins of the doping-induced modulation of the transition temperature are still controversial. In particular it is known that coupling to the lattice plays an important role in driving the transition in the rare earth nickelates [17-19]. However, the concepts discussed here are general and apply to any Mott material exhibiting a metal-insulator transition controlled by band filling.

Figure 1 shows band diagrams of an epitaxial heterostructure that can be used to remotely-dope a Mott material by charge transfer from a highly *n*-doped band insulator. For example, $SrTiO_3$ can be degenerately *n*-doped by La or Nb to achieve carrier concentrations as high as $10^{21}$ cm$^{-3}$ [20,21]. Before contact, as shown in Fig. 1(a), the Mott material (*R*NiO$_3$) is insulating as the operating temperature is *just below* the metal-insulator transition temperature of the undoped material, $T_{MIT}$ (200 K for NdNiO$_3$). The exact band offsets at *R*NiO$_3$/SrTiO$_3$ interfaces are not well known and we assume ~1 eV [22]. In contact, the Fermi levels align [Fig. 1(b)], and electrons transfer into the NdNiO$_3$, depleting electrons in the SrTiO$_3$. If sufficient electrons are transferred, the NdNiO$_3$ near the interface will be metallic. The doping in the SrTiO$_3$ is set back somewhat to mitigate leakage by tunneling when the junction is under bias. Figure 2 shows a schematic of a depletion mode MMFET.

The density of carriers transferred into the Mott material is greatest at the interface. The exponential decrease with distance from the interface can be described using Thomas-Fermi screening, which is characterized by a length, $k_{TF}^{-1}$:

$$k_{TF} = \sqrt{\frac{e(dn/d\mu)}{\varepsilon_{RNO}}} \, , \qquad (2)$$

where $e$ is the electron charge, $\varepsilon_{RNO}$ is the dielectric permittivity of the nickelate (taken to be $30\varepsilon_0$) and $dn/d\mu$ is the density of states provided by the *two* hybridized $e_g$ orbitals per nickel (per unit cell). Assuming a band width for the oxygen-hybridized $e_g$ states of order of 6 eV, the density of states is estimated to be $5.2\times10^{27}$ m$^{-3}$V$^{-1}$. In the metallic phase, the number of electrons transferred into the NdNiO$_3$ with a thickness $t$, is derived as follows:

$$n_{RNO} = \left(\frac{dn}{d\mu}\right)\int_0^t \phi_{RNO}\exp(-k_{TF}z)dz = \sqrt{\left(\frac{dn}{d\mu}\right)\frac{\varepsilon_{RNO}}{e}}\phi_{RNO}\left[1-\exp(-k_{TF}t)\right] \, , \qquad (3)$$

were $\phi_{RNO}$ is the band bending in the $R$NiO$_3$. In the insulating phase, we assume that the density of states is much smaller than in the metallic phase but sufficient to locate the chemical potential near the middle of the Hubbard gap. Assuming no loss of charge to interface traps, the transferred electrons will equal in number those depleted in the La:SrTiO$_3$, $n_{STO} = n_{RNO}$ with:

$$n_{STO} = \sqrt{\frac{2\varepsilon_{STO}N_{La}\phi_{STO}}{e}} \, , \qquad (4)$$

where $N_{La}$ is the La dopant density and $\varepsilon_{STO}$ is the dielectric constant of the SrTiO$_3$, taken as $300\varepsilon_0$. With reference to Figure 1(b), the charge transferred is constrained by potential continuity:

$$\phi_{RNO} + \phi_{STO} + \phi_{SB} + V - V_B = 0, \tag{5}$$

where $V$, $V_B$, and $\phi_{SB}$ are the externally applied bias, the band offset between SrTiO$_3$ and $R$NiO$_3$ under "flat band" condition and the potential change in the undoped setback layer, respectively. The inset in Figure 3(a) shows the ratio of transferred carrier density at the interface ($z = 0$) relative to the carrier density in bulk $R$NiO$_3$ as a function of La donor concentration in SrTiO$_3$ and with a 5 nm undoped setback layer in the SrTiO$_3$. The relative change in interfacial carrier concentration in the $R$NiO$_3$ increases from 2 % to 16 % as the dopant concentration in the SrTiO$_3$ increases from $10^{18}$ cm$^{-3}$ to $10^{21}$ cm$^{-3}$. Given that a few percent of chemical doping lowers $T_{MIT}$ by several tens of degrees in materials such as NdNiO$_3$ [15], $10^{21}$ cm$^{-3}$ should be more than sufficient to induce a transition to the metallic state at the interface, *if the operating temperature is just below $T_{MIT}$*. The density of transferred carriers exponentially decays over a distance dictated by the Thomas-Fermi wave vector. Figure 3(a) shows the spatial distribution of the transferred carriers from $10^{21}$ cm$^{-3}$ La:SrTiO$_3$ as a function of applied voltage in a NdNiO$_3$ film of thickness of ~ 1.5 nm, as given by:

$$n_{RNO}(z,V) = n_{RNO}(0,V)\exp(-k_{TF}z) \tag{6}$$

If we use 1% as a critical concentration of excess electrons for metallic behavior, Fig. 3(a) indicates that the entire channel will be metallic with no bias. As a negative bias is applied to the La:SrTiO$_3$ gate, the number of transferred electrons decreases as does the fraction of the film in the metallic state. The effective channel thickness continues to decrease with gate bias; eventually the excess carrier density is everywhere below 1 %, the $R$NiO$_3$ is insulating and the channel is turned off [see also Fig. 1(c)].

The gate voltage dependent channel sheet conductivity can be calculated by using the known resistivity of NdNiO$_3$ as a function of chemical doping at 150 K [15] (with other $R$NiO$_3$'s, $T_{MIT}$ can be shifted to near room temperature or above [14]). The total sheet channel conductivity, $g_c$, was obtained by summing the two-dimensional conductivity, $\sigma_n t_n$, of layers of thickness $t_n$, using the transferred electron density as a function of position (see Supporting Information):

$$g_c = \sum \sigma_n t_n \qquad (7)$$

From the sheet conductivity, the transistor drain current vs. drain voltage ($I_{SD}$ vs. $V_{SD}$) characteristics can be calculated for $V_{DS} < V_b - V_G$ by the following equation:

$$I_{DS} = \frac{w}{l} \int_0^{V_{DS}} dV \, g_c(V + V_G), \qquad (8)$$

where $l$ is the channel length (taken to be 30 nm in the calculation) and $w$ is the width. Figure 3(b) shows the device characteristics as a function of $V_{SD}$. The device is essentially a depletion mode device. An applied gate voltage removes the excess electrons, returning the Mott material to the insulating ("off") state. Although the device, as configured, is limited for positive source-drain characteristics the transconductance for both positive or negative source drain bias is large, of the order of $g_m = \left(\partial I_{SD}/\partial V_G\right)_{V_{SD}=const}$ ~ 3000 mS/mm at a source-drain voltage of $V_G$ of +/- 0.45 V. This is substantially higher than $g_m$ values of state-of-the-art Si FETs. We also note that the device will also support large current drive, several amperes/mm.

A depletion mode device is perhaps the simplest manifestation of a MMFET. Using a wider band gap insulator in place of the set back layer would open the possibility of enhancement mode operation. A p-channel device would follow if the interfacial region of the MMFET were chemically doped with divalent cations [15]. Finally we note that the MMFET is intrinsically an ambipolar device, which may encourage electronic applications beyond switches [23].

Towards practical realization of the MMFET, critical materials issues must be addressed. Foremost, despite extensive work on chemical doping, which purports to separate the effect of band filling from lattice distortions on the metal insulator transition, experimental demonstration of pure band filling control is limited. The MIT of the rare earth nickelates in particular may be more sensitive to lattice distortions than to a change in band filling. Thus the most important goal towards the realization of any Mott FET should be research that is directed to identify suitable Mott materials that respond to a change in band filling. There are many possible candidate materials in addition to the rare earth nickelates. For example, the rare earth titanates are well known to show metal-insulator transitions that are associated with orders of magnitude change in conductivity upon hole doping [24]. Further, there is little documentation of band offsets in relevant oxide heterostructures. Disorder induced scattering and disorder broadening of the phase transition will compromise any phase change Mott FET. The effects of quantum confinement and surface scattering are largely unexplored. It should be noted, however, that for the MMFET only the interfacial layer plays a role; thicker epilayers would mitigate at least the effects of surface roughness.

**Acknowledgement:**

The authors gratefully acknowledge support by a MURI program of the Army Research Office (grant # W911-NF-09-1-0398) as well as useful discussions with Mark Rodwell.

**Figure Captions**

**Figure 1 (color online)**

Schematic energy band diagrams of a Mott material, such as $R$NiO$_3$, interfaced with a highly-doped band insulator, such as La-doped SrTiO$_3$, (a) before and (b) after contact and (c) after contact with an applied voltage to achieve flat band conditions. The dashed lines in (b) and (c) indicates half-filling of the conduction band of the Mott material, which is assumed to be metallic if the band filling deviates from half-filling by a certain amount (see text). The dotted line indicated the position of the undoped SrTiO$_3$ set-back layer. UHB = upper Hubbard band of the Mott material.

**Figure 2 (color online)**

Schematic of a MMFET.

**Figure 3 (color online)**

(a) Ratio of modulated carrier density relative to the carrier density in the $R$NiO$_3$ as a function of position and applied voltage for a 1.5 nm thick RNiO$_3$ film that is interfaced with a SrTiO$_3$ film doped with $10^{21}$ cm$^{-3}$ of La. The carrier density of bulk RNiO$_3$ ($n_0$) is assumed to be one Ni e$_g$ electron per unit cell, or ~ $1.8 \times 10^{22}$ cm$^{-3}$. The applied voltage is changed in increments on 0.1 V between 0 and 0.9 V. The dashed line indicates an excess carrier density of 1 %. The inset shows of modulated carrier density at the interface as a function of La concentration in the SrTiO$_3$. (b) Current-voltage characteristics of the MMFET calculated from the parameters of chemically doped

NdNiO$_3$, with the gate voltage as parameter. The labels indicate the gate voltage (in Volts), which is varied between 0 and 0.9 V. The transistor characteristics are terminated according to $V_{SD} + V_G < V_B$. Otherwise the Schottky junction in the SrTiO$_3$ will be forward biased.

**Figure 1**

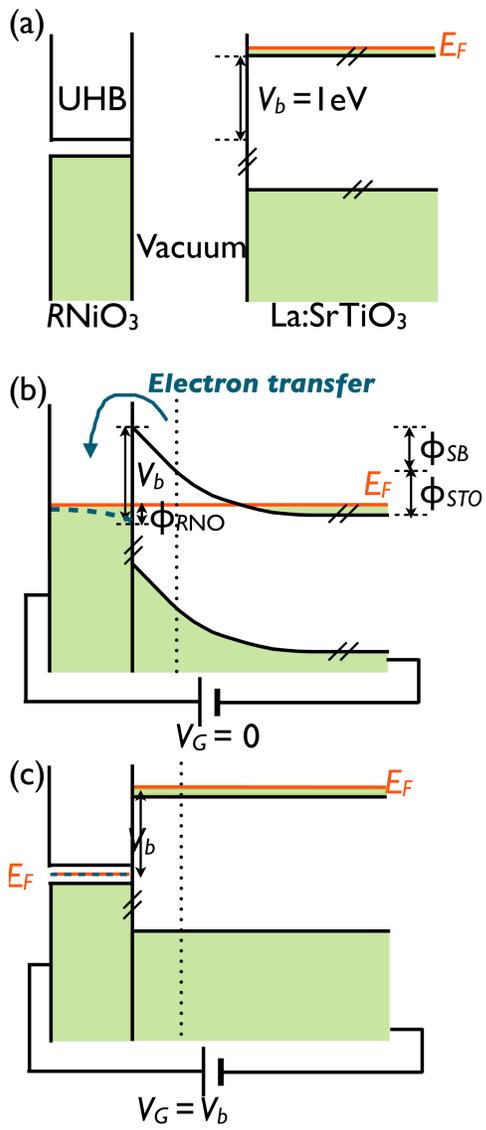

**Figure 2**

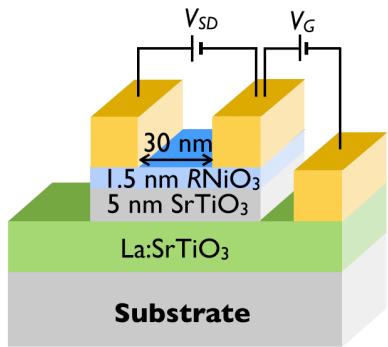

**Figure 3**

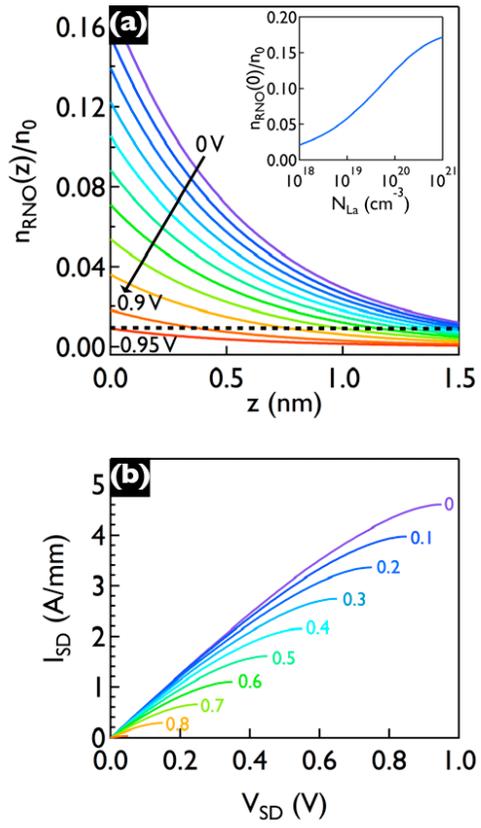

# Supporting information: A heterojunction modulation-doped Mott transistor

**Junwoo Son[1], Siddharth Rajan[2], Susanne Stemmer[1], and S. James Allen[3]**

[1] Materials Department, University of California Santa Barbara, Santa Barbara, CA 93106, U.S.A.
[2] Department of Electrical and Computer Engineering, The Ohio State University, Columbus, OH 43210 USA
[3] Department of Physics, University of California, Santa Barbara, CA 93106-9530, U.S.A.

## 1. Derivation of $f_\tau$ for a MMFET

We use a simple small signal model to derive the unity current gain frequency of the MMFET. In the case of a traditional FET with gate length $l$, width $W$ (model shown in Figure S1), the input (displacement) and output (conduction) currents for a small signal excitation $v_{gs}$ are given by $i_{IN} = j\omega W l C_{gs} v_{gs}$ and $i_{OUT} = W C_{gs} v_{gs} v_{eff}$. Here, $v_{eff}$ is the effective electron velocity of the charges induced in the channel. The charge induced in the channel by the excitation is $C_{gs} v_{gs}$. The ratio of this gives the current gain to be

$$|h_{21}| = \frac{v_{eff}}{\omega l} \qquad (S.1)$$

and the unity current gain (or cutoff frequency) to be

$$f_\tau = \frac{v_{eff}}{2\pi l} . \qquad (S.2)$$

In the case of the MMFET, for the same input excitation, $v_{gs}$, the input current is identical to the equation for the normal FET, $i_{IN} = j\omega W l C_{gs} v_{gs}$. However, since the output charge induced in the channel is higher (due to the Mott transition in the channel layer), the conducting charge modulated in the channel is higher by a factor $K$, and is given by $K C_{gs} v_{gs}$, and the output current is now given by $i_{OUT} = K W C_{gs} v_{gs} v_{eff}$. Taking the ratio $i_{OUT}/i_{IN}$ allows us to estimate the current gain and cutoff frequency for the MMFET:

$$|h_{21}| = \frac{K v_{eff}}{\omega l} \qquad (S.3)$$

$$f_\tau = \frac{K v_{eff}}{2\pi l} \qquad (S.4)$$

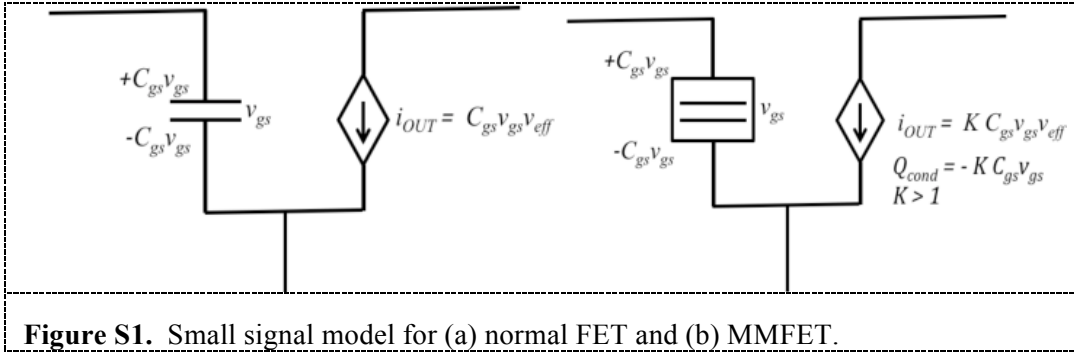
**Figure S1.** Small signal model for (a) normal FET and (b) MMFET.

## 2. Estimate of the channel conductivity of a MMFET with an NdNiO₃ channel

The conductivity (or resistivity) of NdNiO$_3$ as a function of doping concentration was calculated using the bulk resistivity data as a function of temperature and chemical doping from ref. [1]. From the data in ref. [1], $T_{MIT}$ shifts by 50 K per 1 % hole doping although the device being modeled is controlled by electron (not hole) transfer. By shifting the temperature-dependent resistivity plot of undoped NdNiO$_3$ by 5 K per 0.1 % electron doping, the resistivity of electrostatically-doped NdNiO$_3$ at 150 K was obtained in steps of 0.1 %, as shown in Table I.

**Table I:** Resistivity of NdNiO$_3$ as a function of doping at 150 K, estimated from the bulk data in ref. [1].

| Doping (%)       | 0   | 0.1  | 0.2  | 0.3  | 0.4 | 0.5 |
|------------------|-----|------|------|------|-----|-----|
| Resistivity (Ωcm)| 2   | 2    | 1.8  | 1.5  | 1   | 0.6 |

| Doping (%)       | 0.6  | 0.7  | 0.8  | 0.9    | 1      |
|------------------|------|------|------|--------|--------|
| Resistivity (Ωcm)| 0.35 | 0.06 | 0.01 | 0.0015 | 0.0008 |

The channel sheet conductivity was the calculated by slicing the 1.5 nm-thick NdNiO$_3$ film in thickness increments corresponding to 0.1 % changes in doping. Figure S2 shows the spatial distribution of carriers transferred from La:SrTiO$_3$ ($10^{21}$ cm$^{-3}$) to the NdNiO$_3$ under an applied bias of 0.9 V. The sheet conductivity, $g_c$, was calculated from the following relationship:

$$g_c = \frac{z_1}{\rho_{1\%}} + \frac{z_2 - z_1}{\rho_{0.9\%}} + \cdots + \frac{z_8 - z_7}{\rho_{0.3\%}} + \frac{z_9 - z_8}{\rho_{0.2\%}} + \frac{1.5\text{ nm} - z_9}{\rho_{0.1\%}}, \quad (S.5)$$

where $\rho_{n\%}$ is the resistivity of NdNiO$_3$ with $n\%$ doping at 150 K. By carrying out this calculation for different voltages between the channel and doped SrTiO$_3$ the device conductance per mm of gate width for a 30 nm long channel is obtained, as shown in Figure S3. To obtain the channel current-voltage characteristic for source-drain voltages comparable to the gate voltage we recognize that the effective gate voltage depends on the position between the source and drain in the channel. As such we can relate the current density to the applied source-drain and gate voltages, observing the polarity shown in Figure 2:

$$I\frac{l}{w} = \int_0^{V_{SD}} dV\, g_c [V + V_g], \quad (S.6)$$

where $I/w$ is the current density and $l$ is the channel length, taken as 30 nm. Note that $g_c$ is a strong function of position for source-drain voltages comparable to the gate voltage. For small $V_{SD} \ll V_G$ the current density is $I/w = V g_c(V_G)/l$ and the conductance per mm per gate or channel length, is $(I/V)(1/w) = g_c(V_G)(1/l)$ and for $l$ = 30 nm we recover Figure S3. In Fig. 3(b), the transistor characteristics are terminated according to $V_{SD} + V_G < V_B$, where $V_B$ is the heterostructure offset voltage measured from the conduction band edge of the SrTiO$_3$ and the Fermi energy of the NdNiO$_3$. Otherwise the Schottky junction in the SrTiO$_3$ will be forward biased.

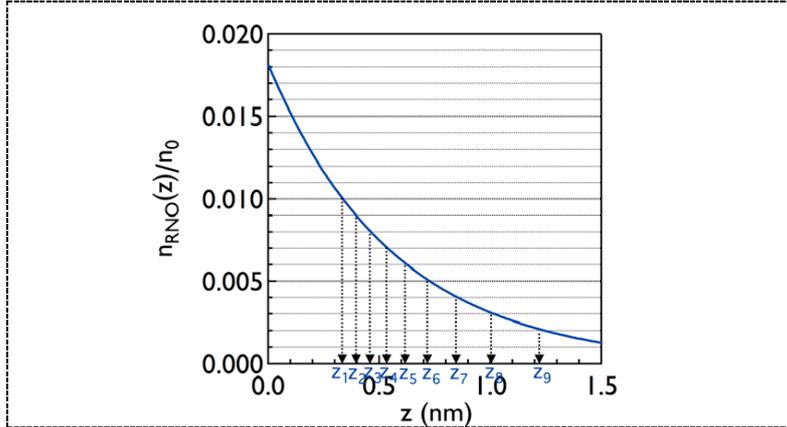

**Figure S2.** Spatial distribution of carriers transferred from La:SrTiO$_3$ ($10^{21}$ cm$^{-3}$) to the NdNiO$_3$ under an applied bias of 0.9 V (blue line) and the thickness increments that were used to calculate the channel sheet conductivity.

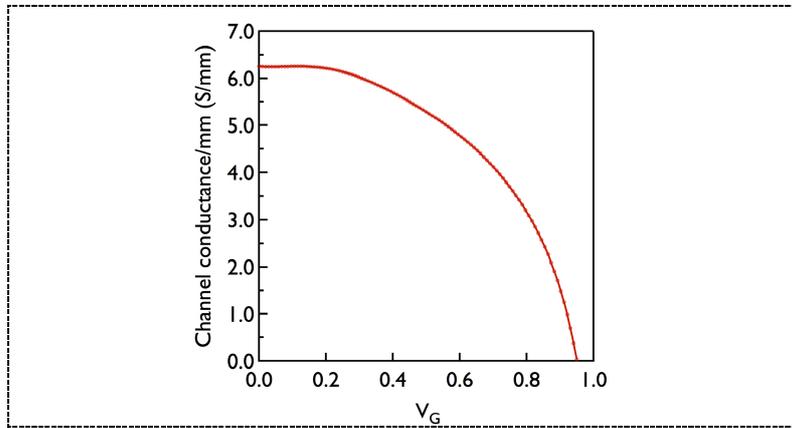

**Figure S3.** Channel conductance/mm width vs. gate voltage for a 30 nm gate or channel length.

## 3. References
[1] J. L. García-Muñoz, M. Suaaidi, M. J. Martínez-Lope, and J. A. Alonso, Phys. Rev. B **52**, 13563 (1995).